# Comparative Study of the Function Overloading and Function Overriding Using C++

Dr. Brijender Kahanwal
Computer Science & Engineering
Galaxy Global Group of Institutions,
Dinarpur, Ambala, India
imkahanwal@gmail.com

*Abstract*—**In the Object-Oriented Programming Systems (OOPS), these two concepts namely function overloading and function overriding are a bit confusing to the programmers. In this article this confusion is tried to be removed. Both of these are the concepts which come under the polymorphism (poly means many and morph mean forms). In the article the comparison is done in between them. For the new programmers and the learners, it is important to understand them. The function overloading [1] is achieved at the time of the compile and the function overriding is achieved at the run time. The function overriding always takes place in inheritance, but the function overloading can also take place without inheritance.**

## I. INTRODUCTION

In software development, object oriented paradigm plays an important role for the novel issues of the computing like concurrent, distributed, and parallel etc. There are many concepts of these systems like object, class, data abstraction (data hiding), polymorphism, inheritance, generic programming and exceptional handling. The concept polymorphism is most interesting to the developers. It has so many sub concepts in it like function overloading, virtual function, operator overloading, and function overriding. Here function overloading and function overriding are explored. In the next section function overloading is described with the help of the figures as well as program implementation. And the next to next section the function overriding is explored with the help of the implementation examples.

## II. FUNCTION OVERLOADING

It is the static type of polymorphism in which the function or method name are same, but the number of arguments or the types of the arguments or order of arguments may vary among them [1]. It may occur in the same class or in inheritance or without the inheritance or in template-based functions and other normal functions or in between constructors of the same class or scope base function overloading 1].

As described in the Figure 1, there are two classes A (Base class) and B (Derived class) and they have methods get ( ) and display ( ) for class A and get ( ), get (int) methods for class B. As the concept of the inheritance, the base class methods get ( ) and display ( ) are inherited in the class B and the class B also have its own methods get ( ) and get (int) which has been overloaded and both these overrides the base class method get ( ). After the inheritance, the class B has four methods in it two methods are its own and two methods are from the class A. Here the derived class B's object can approach directly the three methods namely B's get ( ) and get (int) and A's display ( ). The implementation is shown in the Figure 2.

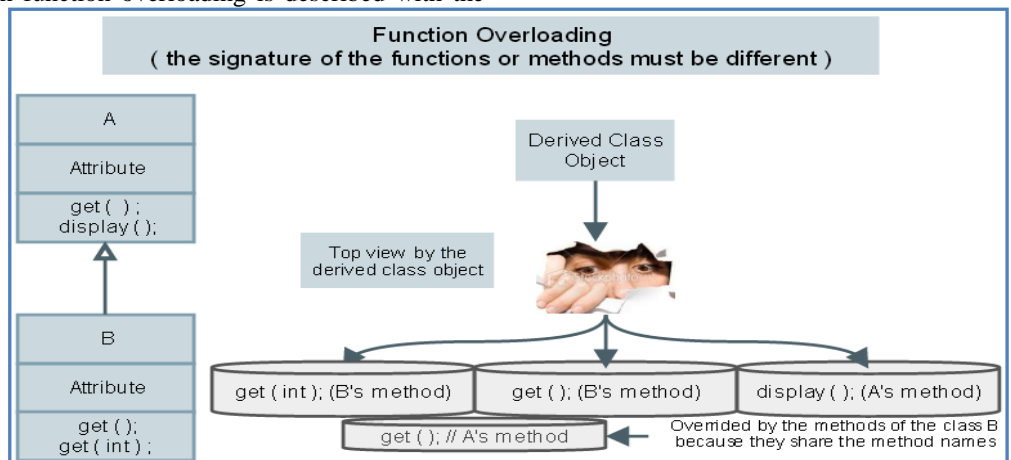

Figure 1. Function overloading in the class B of get function or method

```
#include <iostream.h>
class A
{
 public:
           get ( );
           display ( );
};
A:: get ( )
{
 cout <<"\n A class member function\n";
 return (0);
}
A :: display ( )
{
 cout << "\n A class display method \n";
 return (0);
}
class B : public A
{
 private:
           int b;
 public:
           get ( int x);
           get ( );
};
B:: get ( int x)
{
 b=x;
 cout<< "\nget with one argument function of the class B \n";
 return (0);
}
B::get ( )
{
  cout<<"\nI am the get ( ) method of class B\n";
  return (0);
}
void main ( )
{
 int p;
 cout<< "\nEnter an integer \n";
 cin>> p;
 B b1; // object b1 is declared for the derived class
 b1.get ( );  //Due to overriding, the B's get ( ) method is called
 b1.A::get ( ); // to call the class A's method here.
 B1.get ( p ); // overloaded method of class B is called
 b1.display ( ); // class A's method is called here
}
```

| OUTPUT |
|---|
| Enter an integer |
| 32 |
| I am the get ( ) method of class B |
| A class member function |
| get with one argument function of the class B |
| A class display method |

Figure 2. Function overloading implementation coding in inheritance of classes A and B.

### III. FUNCTION OVERRIDING

It permits to substitute the inherited method with a novel implementation with the same function or method names. In this concept, signature may or may not be the same, the function or method name must be same. In the Figure 3, there are four classes class D inherits the class C and class B inherits the class A involved in two different inheritances. In the case of classes C and D, both the classes have their own independent methods with the same signatures namely get (int). The derived class D's method overrides the base class C's method get (int). But in another case of classes A and B, both the classes have their own methods get ( ) and get (int) respectively. Here the signatures of the functions or methods are different one has no argument and another has one argument, but they share the function names. Here the function get (int) of derived class B overrides the function get ( ) of base class A. Both of these are also explored with the help of individual programs in Figures 4.a and 4.b.

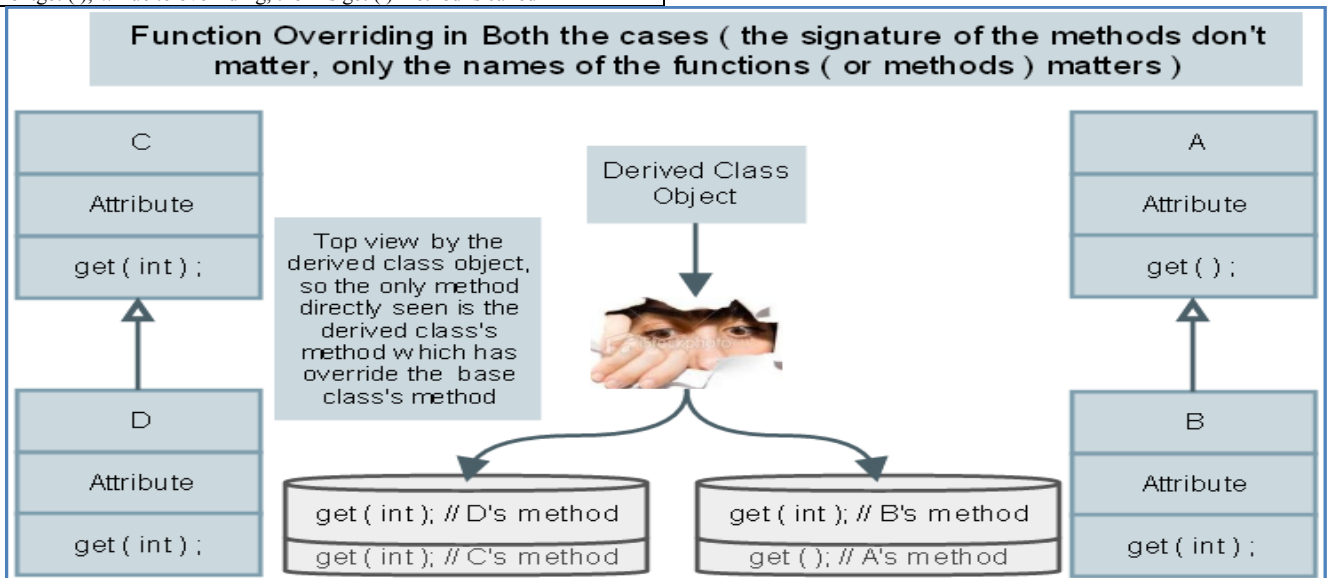

Figure 3. Function overriding in inheritance with the help of the classes C, D and A, B.

```
#include <iostream.h>
class C
{
 private:
            int a;
 public:
            get ( int );
            put ( );
};
C:: get ( int x)
{
 a=x;
 return (0);
}
C :: put ( )
{
 cout << "The value of the class C's object =" << a;
return (0);
}
class D : public C
{
 private:
            int b;
 public:
            get ( int x);
            display ( );
};
D:: get ( int x)
{
 b=x;
 return (0);
}
D :: display ( )
{
 cout << "\nThe value of the class D's object="<< b;
 return (0);
}
void main ( )
{
 int p, q;
 cout << "Enter two integer \n";
 cin >> p >> q;
 D d1; // object d1 is declared for the derived class
 d1.get (p);
// class D's method is called here because of overriding
 d1.put ( ); // class C's method called
 d1.display ( ); // class D's method called
 d1.get ( q ); // again overridden method (D's) is called
 d1.display ( ); // class D's method is called
 d1.C::get ( p);
// overriding is resolved with the help of the scope resolution
//operator to call the class C method
 d1.put ( );
}
```

**OUTPUT**

Enter an integer
32
42
The value of the class C's object =26376
The value of the class D's object=32
 The value of the class D's object=42
The value of the class C's object =32

Figure 4.a Function overriding implementation coding with the help of classes C and D with same signatures.

```
#include <iostream.h>
class A
{
 public:
            get ( );
};
A:: get ( )
{
 cout<<"\n Class A's get () method is called\n";
 return (0);
}
class B : public A
{
 private:
            int b;
 public:
            get ( int x);
};
B:: get ( int x)
{
 b=x;
 cout << "\n Class B's get (int) method is called\n";
 return (0);
}
void main ( )
{
 int p;
 cout << "\nEnter an integer \n";
 cin >> p;
 B b1; // object b1 is declared for the derived class
 //b1.get ();
 // class B's method is called here because of overriding
 // and the compiler will show the error message as follows
 //"too few parameters in call to 'B::get (int)'"
//so it will not work and I made it as a comment
 b1.get (p); // class B's method called
 b1.A::get ( ); // now the class A's method is called
//with the help of the scope resolution operator
}
```

**OUTPUT**

Enter an integer
12
 Class B's get (int) method is called
 Class A's get ( ) method is called

Figure 5.b Function overriding implementation coding with the help of classes A and B with different signatures.

IV. COMPARISON BETWEEN FUNCTION OVERLOADING AND FUNCTION OVERRIDING

Function overloading and function overriding have a lot of difference in between them, but they sound a bit similar for the innovative learners. The comparison of them is well described with the help of the Table 1.

Table 1. Comparison between the function overloading and function overriding

| SR. NO. | FUNCTION (METHOD) OVERLOADING | FUNCTION (METHOD) OVERRIDING |
|---|---|---|
| 1 | It may occur without inheritance. | It always requires the inheritance to occur. |
| 2 | In it functions or methods share only the names of the functions or methods. **(Different signatures)** | In it the methods of the classes share the name of the function only not the complete signatures. **(Different or same signatures)** |

| 3 | It takes place at compile time. | It takes place at the runtime. |
|---|---|---|
| 4 | It is the form of compile time (static) polymorphism. | It is the form of runtime (dynamic) polymorphism. |
| 5 | It requires at least one class to occur. | It requires at least two classes (one base class and another derived class) to occur. |
| 6 | It takes place in between the functions or methods which are in the same scopes or different scopes. | It takes place in between the methods which are in the different scopes only. |
| 7 | All methods or functions will be called directly (without the help of scope resolution operator) because there is no overriding. | The derived class methods override the base class methods which share the same names (if you want to access the base class function or method then we has to use scope resolution operator). |

## V. Implementation

All the execution of the C++ programs has been done on the machine Intel ® Core ™ 2 Duo CPU T6500 @ 2.10 GHz and the primary memory is 2 GB RAM. The operating system was Microsoft Windows XP ProfessionalVersion 2002 Service Pack 2. All the programs are also run on the freely available online compiler C++ 4.7.2 (gcc-4.7.2) on ideone.com. It is an online compiler and debugging tool which allows us to compile and run code online in more than 40 programming languages. The web browser, Google Chrome was used to access the website www.ideone.com.

## VI. Conclusion

For developing the robust softwares, it is good to have a sound understanding of the programming language. Regarding a good understanding and evaluation of object-oriented concepts, function overloading and function overriding, this article enriches the knowledge of the C++ lovers. Here so many variations are shown about the functions overriding with the help of suitable examples with their implementations and results. The developers will take an advantage after go through it. I hope it will be a good diet for them.

## Acknowledgment

The author is greatly thanks to the unspecified reviewers especially Mr. Deepak Kumar who have commented on this creation and because of them it has been finalized.